\documentclass[aps,pra,reprint,twocolumn,showpacs,floatfix,superscriptaddress]{revtex4-1}

\usepackage{amssymb,amsmath,amstext}                %%   American Physical Society math etc extensions
\usepackage{graphicx}                                               %%   Include figure files                                            
\usepackage{epstopdf}                                               %%   help with eps -> pdf 
\usepackage{color}                                                     %%   change color
\usepackage{bm}                                                        %%   bold math                                    
\usepackage{appendix}                                              %%   formating appendicies
\usepackage[utf8]{inputenc}
\usepackage{bbold}
\usepackage{bbm}
\usepackage{ulem}
\usepackage{latexsym}
\usepackage{xcolor}
\usepackage{braket}
\definecolor{lblue} {RGB}{51,71,158}
\usepackage[colorlinks=true,citecolor=blue,linkcolor=blue,urlcolor=lblue]{hyperref}

\newcommand{\be}{\begin{equation}}
\newcommand{\ee}{\end{equation}}

\begin{document}

%%%% Article title to be placed here
\title{Cold atoms meet lattice gauge theory}

\author{%%%% Author details
Monika Aidelsburger$^{1,2}$, Luca Barbiero$^3$, Alejandro Bermudez$^4$, Titas Chanda$^5$, Alexandre Dauphin$^3$, Daniel González-Cuadra$^3$, Przemysław R. Grzybowski$^6$, Simon Hands$^7$, Fred Jendrzejewski$^{8}$, Johannes J\"unemann$^{9}$, Gediminas Juzeliunas$^{10}$, Valentin Kasper$^3$, Angelo Piga$^3$, Shi-Ju Ran$^{11}$, Matteo Rizzi$^{12,13}$, G\'erman Sierra$^{14}$, Luca Tagliacozzo$^{15}$, Emanuele Tirrito$^{16}$, Torsten V. Zache$^{17,18}$, Jakub Zakrzewski$^{5}$, Erez Zohar$^{19}$, Maciej Lewenstein$^{3,20}$
}

%%%%%%%%% Insert author address here
\address{$ $Fakult\"at f\"ur Physik, Ludwig-Maximilians-Universit\"at M\"unchen, 80799 Munich, Germany\\
$^{2}$ Munich Center for Quantum Science and Technology (MCQST), 80799 M\"unchen, Germany       \\
$^3$ ICFO - Institut de Ciencies Fotoniques, The Barcelona Institute of Science and Technology,
 08860 Castelldefels (Barcelona), Spain\\
$^4$ Departamento de F\'isica Teorica, Universidad Complutense, 28040 Madrid, Spain \\
$^5$Institute of Theoretical Physics, Jagiellonian University in Krak\'ow,  30-348 Krak\'ow, Poland \\
$^6$ Faculty of Physics, Adam Mickiewicz University in Pozna\'n, Pozna\'n, Poland\\
$^7$Department of Physics, Faculty of Science and Engineering,
Swansea University, Swansea SA28PP, United Kingdom\\ 
$^8$Universit\"at Heidelberg, Kirchhoff-Institut f\"ur Physik,
69120 Heidelberg, Germany\\
$^9$Johannes Gutenberg-Universit\"at, Institut f\"ur Physik, 55128 Mainz, Germany\\
$^{10}$ Institute of Theoretical Physics and Astronomy, Vilnius University,  LT-10257 Vilnius, Lithuania\\
$^{11}$Department of Physics, Capital Normal University, 100048 Beijing, China \\
$^{12}$Forschungszentrum J\"ulich, Institute of Quantum Control, Peter Grünberg Institut (PGI-8), 52425 J\"ulich, Germany\\
$^{13}$Institute for Theoretical Physics, University of Cologne, 50937 K\"oln, Germany\\
$^{14}$Instituto de F\'isica Te\'orica, UAM/CSIC, Universidad Aut\`onoma de Madrid, Madrid, Spain\\
$^{15}$Departament de Física Qu\`antica i Astrofísica and Institut de Ci\`encies del Cosmos (ICCUB),
Universitat de Barcelona, 08028 Barcelona, Catalonia, Spain\\
$^{16}$International School for Advanced Studies (SISSA), 34136 Trieste, Italy\\
$^{17}$Center for Quantum Physics, University of Innsbruck, 6020 Innsbruck, Austria\\
$^{18}$Institute for Quantum Optics and Quantum Information of the Austrian Academy of Sciences, 6020 Innsbruck, Austria\\
$^{19}$Racah Institute of Physics, The Hebrew University of Jerusalem, Jerusalem 91904,  Israel\\
$^{20}$ICREA, Passeig Lluis Companys 23,  08010 Barcelona, Spain
%\end{mdframed}
}

\begin{abstract}
\noindent The central idea of this review is  to consider quantum field theory models relevant for particle physics and replace the fermionic matter in these models by a bosonic one. This is mostly motivated by the fact that bosons are more ``accessible'' and easier to manipulate for experimentalists, but this ``substitution'' also leads to new physics and novel phenomena. It allows us to gain new information about among other things confinement and the dynamics of the deconfinement transition. We will thus consider bosons in dynamical lattices corresponding to the bosonic Schwinger or Z$_2$ Bose-Hubbard models. Another central idea of this review concerns atomic simulators of paradigmatic models of particle physics theory such as the Creutz-Hubbard ladder, or Gross-Neveu-Wilson and Wilson-Hubbard models. Finally, we will briefly describe our efforts to design experimentally friendly simulators of these and other models relevant for particle physics.
\end{abstract}
%%%%%%%%%%%%%%%%%%%%%%%%%%%
\maketitle
%%%%%%%%%% Insert the texts which can accomdate on firstpage in the tag "fmtext" %%%%%

\section{Introduction}
Quantum simulators (QS) \cite{trabesinger2012quantum} constitute one of the pillars of quantum technology \cite{Acin2018}. Although quantum advantage with QS was achieved many years ago \cite{bernien2017probing}, and keeps being repeated in various systems and contexts \cite{Tan2021}, most of the applications of QS concern quantum many-body physics.  QS involve many platforms \cite{Cirac2012}: superconducting circuits \cite{Houck2012}, through ultracold atoms \cite{Bloch2012}, trapped ions\cite{Blatt2012}, 	Rydberg atoms \cite{celi2019emerging}, circuit QED\cite{cqednaturespecial}, photonic systems\cite{Aspuru-Guzik2012}, and more. QS has enriched our understanding of quantum many-body systems in the last decades enormously: from the physics of Fermi-Hubbard models \cite{Dutta2015} to non-trivial aspects of interacting disordered systems and many-body localization\cite{Abanin2019}.

Typically, QS are thought to mimic condensed matter physics, bur as early as 2005 people started to talk about simulations of high energy physics and, in particular, lattice gauge theories (LGT) \cite{buchler2005atomic,Osterloh2005}. Several proposals/designs were later formulated, employing typically quantum link models \cite{Wiese2013} in which gauge fields are represented on the links of the lattice in a  finite-dimensional Hilbert space (see for instance \cite{zohar2011confinement,zohar2012simulating,banerjee2012atomic,banerjee2013atomic,tagliacozzo2013simulation,tagliacozzo2013optical}. These designs can often be analysed efficiently in terms of tensor networks methods (TN) - a European collaboration programme QTFLAG made essential progress in this respect (for a review see \cite{Banuls2020}). Amazingly, many of these theoretical designs were reformulated in an "experiment-friendly" manner, and have found experimental realizations, or at least ``first-experimental-steps-toward-realization'' \cite{martinez2016real,
Schweizer2019,Goerg2019, Mil2020,Yang2020}.

As an example of such LGT design consider a toolbox for LGTs \cite{dutta2017toolbox} where a class of simple two-dimensional models admitting a low energy description in terms of an Abelian gauge theory is proposed. The models display rich phase diagrams, with exotic deconfined phases and gapless phases - a rare situation for 2D Abelian gauge theories emerging from the  additional symmetry in these models. Shaken  ultracold bosonic atoms in optical lattices provide a possible experimental platform for this toolbox proposal.

The present paper for the special issue of the Proceedings of the Royal Society reviews activities in the field of QS of LGT and related models performed by the Quantum Optics Theory group at ICFO in collaboration with many others. It has a review character, but includes also some novel, unpublished results. 

Section~\ref{Titas} presents recent achievements in the physics of the bosonic Schwinger model. Characteristically for cold atoms, experiments are much easier for bosons motivating this work which parallels the more extensively studied fermionic case. Section~\ref{Daniel} describes bosons on lattices in a tight-binding approach via the  Bose-Hubbard model, but with tunnellings mediated by other atomic species with certain symmetries.
Section~\ref{Emanuele} deals with Creutz-Hubbard and Creutz-Ising ladders as well as the paradigmatic Gross-Neveu-Wilson and Wilson-Hubbard models. Last but not least, Section~\ref{Valentin} discusses dynamical gauge field simulators that are the most experimentally friendly. We discuss future perspectives and conclude in Section~\ref{Matteo}.

\section{Bosonic Schwinger model }
\label{Titas}

The renowned fermionic Schwinger model (FSM)  \cite{schwinger_pr_1951, schwinger_pr_1962} describes one of the simplest gauge theories in 1+1 dimensions (1D): it has been studied extensively over the years  \cite{coleman_aop_1976, hamer_npb_1982, byrnes_prd_2002, banuls_jhep_2013, buyens_prl_2014}
as it
enjoys interesting features similar to quantum chromodynamics (QCD) in 3+1 dimensions, such as a spontaneous breaking of chiral symmetry and a confinement. 
In this section, we consider its far less-explored bosonic counterpart, the bosonic Schwinger model (BSM) \cite{chanda_prl_2020} that also manifests strong confinement phenomena.

\begin{figure}[t]
  \centering
  \includegraphics[width=0.9\linewidth]{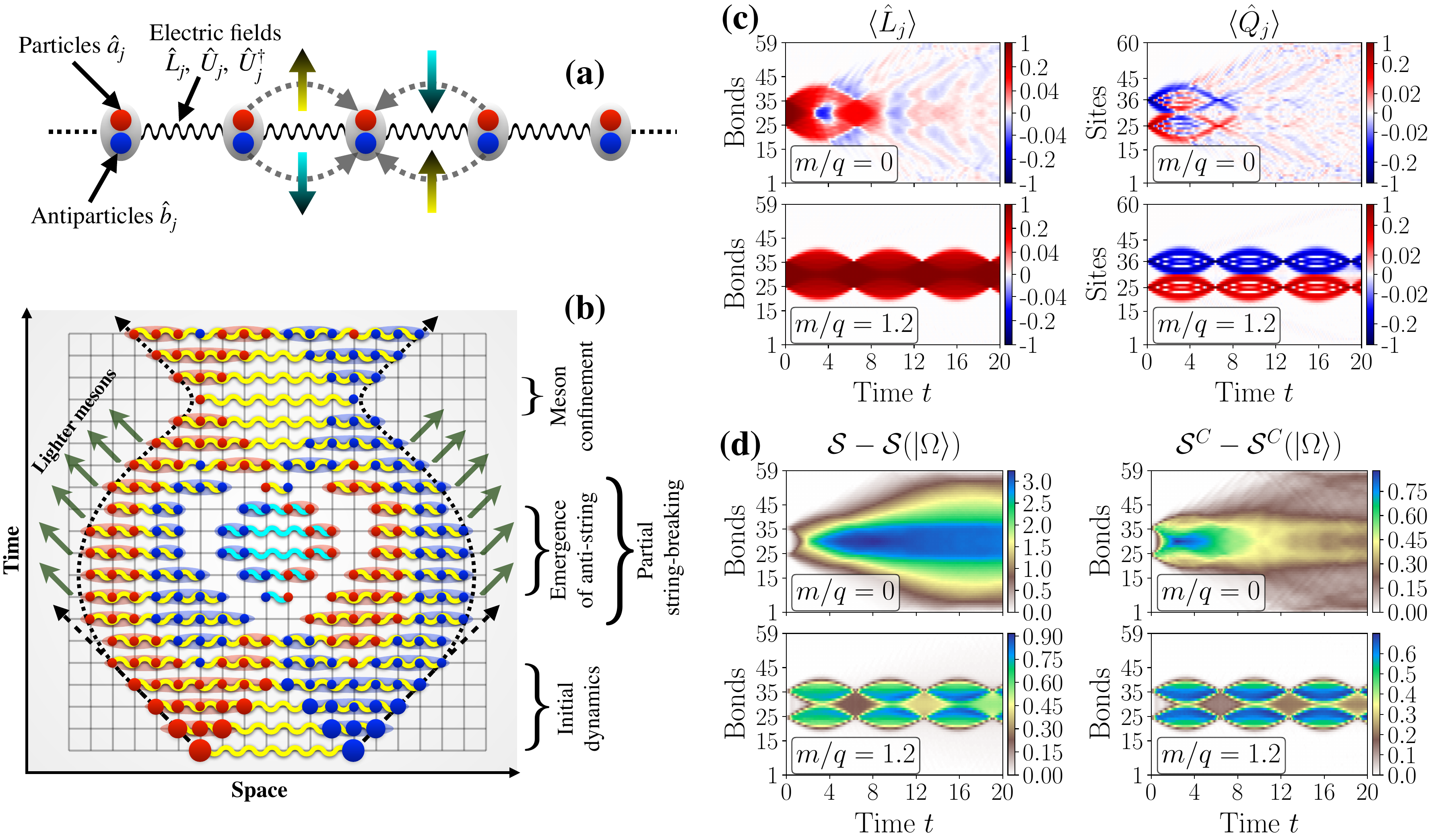}
\caption{\label{fig:bsm} {\bf The bosonic Schwinger model and its time-evolution} \cite{chanda_prl_2020}{\bf :} 
\textbf{(a)} Schematic depiction of BSM, where lattice sites are populated by particles (red circles) and antiparticles (blue circles) and the bonds between neighboring sites hold U(1) electric gauge fields. Left moving particles (antiparticles) raise (lower) the quantum state of the electric field in a corresponding bond, while the opposite holds for right moving bosons.
\textbf{(b)}
Sketch of the confining dynamics of BSM. The system is driven out of equilibrium by creating spatially separated particle-antiparticle pair connected by a string of electric field (a yellow wiggly line).  The strong confinement of bosons bends the trajectory of both excitations. New dynamical charges are created during the evolution that partially screen the electric field. However, the electric field oscillates coherently and may form an anti-string (cyan wiggly line), creating a central core of strongly correlated bosons that is very different from an equilibrium state.
This strange central region survives despite the fact that the boson density in the central region may be depleted through the radiation of lighter mesons that can propagate freely.
\textbf{(c)} Dynamics of the electric field $\langle \hat{L}_j \rangle$ and the dynamical charge $\langle \hat{Q}_j \rangle$, and \textbf{(d)} the same for the total entanglement entropy $\mathcal{S}$ measured across different bonds and its classical part $\mathcal{S}^C$.}
\end{figure}

The BSM is described by relativistic scalar bosons coupled to an U(1) gauge field in 1D. The discretized Hamiltonian of the BSM is given by (see \cite{chanda_prl_2020} for details)
\begin{align}
\hat{H}&= \sum_j \hat{L}_j^2 + 
2 \left(x \left(\left(m/q\right)^2 + 2 x\right)\right)^{1/2} \ \sum_j \big(\hat{a}_j^{\dagger}\hat{a}_j + \hat{b}_j\hat{b}_j^{\dagger}\big) \nonumber \\
&- \frac{x^{3/2}} {\left(\left(m/q\right)^2 +2 x\right)^{1/2}} \sum_j \left[ \big(\hat{a}^{\dagger}_{j+1} + \hat{b}_{j+1}\big) \hat{U}_j \big(\hat{a}_{j} + \hat{b}^{\dagger}_{j}\big) + \text{H.c.}\right],
\label{eq:Hamil_BSM}
\end{align} 
where $q$ is the electric charge, $m$ the bare mass of particles, $x$ is related to the discrete lattice-spacing $a$ by the relation $x = 1/(a^2 q^2)$, $\{\hat{a}_j^{\dagger}, \hat{a}_j\}$, $\{\hat{b}_j^{\dagger}, \hat{b}_j\}$ are bosonic creation-annihilation operators corresponding to particles and antiparticles respectively, and $\hat{L}_j$ the electric field operator residing on the bond between sites $j$ and $j+1$  with $\{\hat{U}_j, \hat{U}_j^{\dagger}\}$ being 
U(1) ladder operators satisfying $[\hat{L}_j, \hat{U}_l] = -\hat{U}_j \delta_{jl}$ and $[\hat{L}_j, \hat{U}^{\dagger}_l] = \hat{U}^{\dagger}_j \delta_{jl}$.
The system is presented schematically in Fig.\ref{fig:bsm}\textbf{(a)}.

The Hamiltonian is invariant under local U(1) transformations: $\hat{a}_j \rightarrow e^{i \alpha_j} \ \hat{a}_j$,
$\hat{b}_j \rightarrow e^{-i \alpha_j} \ \hat{b}_j$,
$\hat{U}_j \rightarrow e^{-i \alpha_j} \ \hat{U}_j \ e^{i \alpha_{j+1}}$, where the corresponding Gauss law generators are given by
$\hat{G}_j = \hat{L}_j - \hat{L}_{j-1} - \hat{Q}_j$,
with  $\hat{Q}_j = \hat{a}_j^{\dagger} \hat{a}_j - \hat{b}_j^{\dagger} \hat{b}_j$ being the dynamical charge. 
We consider the physical subspace spanned by the set of states, $|\Psi\rangle$ that are annihilated by $\hat{G}_j$. 
The low energy spectrum of this system is always gapped even for massless bosons, and as a result the bosons are always confined \cite{chanda_prl_2020}.
As an example  we review, here, the strong confining dynamics of the system of finite size $N$ using matrix-product-states (MPS) techniques \cite{schollwock_aop_2011, paeckel_aop_2019, ran_tensor_2020}. 

The out-of-equilibrium dynamics (OED) of the BSM is initiated by creating two extra dynamical charges of opposite signs, at positions $N/2 - R$ and $N/2+R+1$ respectively, connected by a string of electric field on top of the ground state $|\Omega\rangle$ by the non-local operator
$\hat{M}_R \equiv \left(\hat{a}_{\frac{N}{2} - R}^{\dagger} + \hat{b}_{\frac{N}{2} - R}\right) \left[\prod_{j = \frac{N}{2} - R}^{\frac{N}{2} + R} \hat{U}_{j}^{\dagger}\right]  \left(\hat{a}_{\frac{N}{2} + R + 1} + \hat{b}_{\frac{N}{2} + R +1}^{\dagger}\right)$. 
In an ergodic system, two such excess particles would rapidly delocalize and the system would return to a state indistinguishable from equilibrium. Here we observe something very different as described in the cartoon of the dynamics in Fig.~\ref{fig:bsm}\textbf{(b)}: $(1)$ the light cone of the excitations always bends, representing a slowing down and inversion of their trajectories, as a result of the strong confinement; $(2)$  the initial extended meson formed by the two charges and the electric-flux string connecting them is very robust and the string of electric field joining the two excitations does not break, but rather undergoes at least a couple of coherent oscillations; $(3)$ 
for lighter masses, we observe a string-inversion phenomenon and radiation of lighter mesons from the central region; $(4)$ even once the radiated mesons are free to escape from the confined region and fly away with a constant velocity, as expected in ergodic systems, they leave behind a strongly correlated central core where bosons are confined.
The footprints of these phenomena are presented in Fig.~\ref{fig:bsm}\textbf{(c)} by considering spatio-temporal profiles of the electric field $\langle \hat{L}_j \rangle$ and the dynamical charge $\langle \hat{Q}_j \rangle$ for two values of bare mass $m/q=0$ and $1.2$, that together display all the phenomena listed above. For the massless case, due to a meson radiation from the edges, the confined core of bosons is gradually depleted and disappears after a finite time ($t \approx 10$), while for $m/q=1.2$ both gauge and particle sectors show long-lived coherent oscillations.

In order to better understand the nature of the ergodicity violation, we consider the dynamics of bond-resolved entanglement entropy $\mathcal{S}_j$, measured across the bond between the sites $j$ and $j+1$, 
presented in the left column of Fig.~\ref{fig:bsm}\textbf{(d)}. For  $m/q=0$ most of the entanglement is contained in the central confined region and persists even long after the concentration of bosons in the bulk disappears at around $t \approx 10$, showing a strong memory effect and hence a lack of thermalization. As a result of the  U(1) symmetry related to the conservation of the total charge $\sum_j \hat{Q}_j$, the reduced density matrices of the system are block-diagonal.
There are two contributions to the entanglement entropy \cite{schuch_prl_2004, schuch_pra_2004} 
$\mathcal{S}= \mathcal{S}^C + \mathcal{S}^Q$, a classical part (the Shannon entropy between different quantum sectors), and a quantum part. While the  quantum part $\mathcal{S}^Q$ qualitatively follows the pattern of total entropy $\mathcal{S}$, the classical part $\mathcal{S}^C$ shows strong non-ergodic behavior by sharply demarcating confined and deconfined domains as depicted in 
the right column of Fig.~\ref{fig:bsm}\textbf{(d)}.
Summarizing, the strong spatial inhomogeneity of the time-evolved entropies, especially of their classical part, gives us a clear generic signature of the persistent memory effect and lack-of-thermalization and ergodicity in the system. 

For detailed analysis of such non-ergodic behavior in BSM by means of entropy scaling, we refer the reader to \cite{chanda_prl_2020}. Specifically, it was shown that the central confined region remains non-thermal, while the external deconfined region seems to thermalize -- thus producing  exotic asymptotic states.

\section{Strongly-correlated bosons on a dynamical lattice}
\label{Daniel}

Fermionic matter coupled to dynamical gauge fields has been more extensively explored than its bosonic counterpart, mainly due to the fermionic nature of the matter sector in the Standard Model of particle physics. As presented in the last section, much of the relevant high-energy phenomenology is also present in LGT with bosonic matter, additionally showing novel strongly-correlated effects. These could be further explored using ultracold atoms in optical lattices, with the additional benefit that bosonic atoms are easier to control experimentally. In Sec.~\ref{Valentin} we will present realistic experimental proposals as well as review the current experimental status to simulate LGT with cold atoms. 

\begin{figure}[t]
  \centering
  \includegraphics[width=1.0\linewidth]{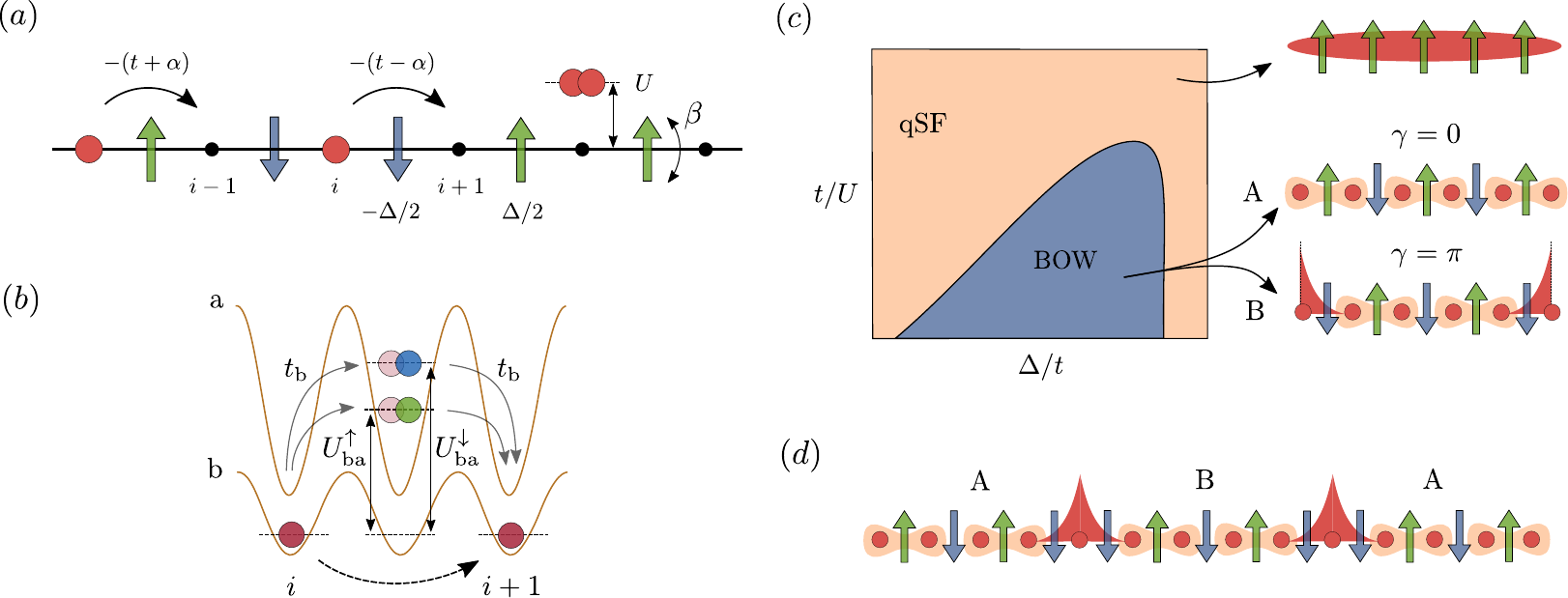}
\caption{\label{fig:scheme} {\bf Symmetry-breaking topological insulators~\cite{Gonzalez-Cuadra_2020a}:} \textbf{(a)} Sketch of the $\mathbb{Z}_2$ BHM~\eqref{eq:z2_bhm}. Bosonic particles (red spheres) hop on a one-dimensional lattice interacting among then and with $\mathbb{Z}_2$ fields (arrows). The latter are located on lattice links and their configuration modifies the tunnelling strength. \textbf{(b)} The model describes a mixture of ultracold bosonic atoms in an optical lattice, where two hyperfine states of one deeply trapped species (green/blue spheres) simulates the $\mathbb{Z}_2$ field~\cite{Gonzalez-Cuadra_2018}. The correlated tunnelling term can be obtained as a second-order density-dependent tunnelling process of the other species~\cite{Chanda_2020}. \textbf{(c)} Qualitative phase diagram at half filling. For strong enough Hubbard interactions, the system undergoes a bosonic Peierls transition from a qSF to a BOW phase where the field orders anti-ferromagnetically~\cite{Gonzalez-Cuadra_2019a}. The two degenerate symmetry-broken patterns (A and B) give rise to insulating states, one of which manifests non-trivial topological properties such as localized edge states with a fractional bosonic number. \textbf{(d)} The presence of dynamical fields and the interplay between symmetry breaking and topological symmetry protection gives rise to strongly-correlated effects that are absent in an static lattice. In the figure, topological defects are shown between the different symmetry-broken field patterns, hosting fractional bosonic states that can move along the system's bulk~\cite{Gonzalez-Cuadra_2020a, Gonzalez-Cuadra_2020b}.}
\end{figure}

In this section, we will explore a further simplification that also leads to interesting phenomena in related models. We start from the simplest LGT with dynamical matter, ie. a one-dimensional chain of bosons coupled to a $\mathbb{Z}_2$ gauge field~\cite{Borla_2020},
\begin{equation}
\label{eq:z2_hamiltonian}
H_{\mathsf{Z}_2}=-\alpha\sum_i\left({b}^\dagger_i{\sigma}^{z\vphantom{\dagger}}_{i,i+1}{b}^{\vphantom{\dagger}}_{i+1}+\text{H.c.}\right)+\beta\sum_i{\sigma}^x_{i,i+1},
\end{equation}
Although a quantum simulation of the full 1D chain is still lacking, the minimal building block of this model, a gauge-invariant correlated tunnelling term, has already been implemented with ultracold atoms using Floquet engineering~\cite{Schweizer2019}. We now consider adding extra terms to the Hamiltonian that break gauge invariance. In particular, we include the standard Bose-Hubbard Hamiltonian describing ultracold bosonic atoms in optical lattices~\cite{Jaksch_1998},
\begin{equation}
\label{eq:bose-hubbard}
H_{\mathsf{BH}}=-t\sum_i \left({b}^\dagger_i{b}^{\vphantom{\dagger}}_{i+1}+\text{H.c.}\right)+\frac{U}{2}\sum_i {n}_i({n}_i-1),
\end{equation}
as well as a parallel field for the gauge degrees of freedom, $H_{\Delta}=\frac{\Delta}{2}\sum_i{\sigma}^z_{i,i+1}$. We denote the full Hamiltonian as the $\mathbb{Z}_2$ Bose-Hubbard model ($\mathbb{Z}_2$ BHM)~\cite{Gonzalez-Cuadra_2018};
\begin{equation}
\label{eq:z2_bhm}
H_{\mathsf{Z_2BH}}=H_{\mathsf{Z}_2}+H_{\mathsf{BH}}+H_{\Delta},
\end{equation}
describing a chain of strongly-correlated bosons whose tunnelling elements depend locally on the $\mathbb{Z}_2$ field configuration (Fig.~\ref{fig:scheme}{\bf (a)}).

The $\mathbb{Z}_2$ BHM could be implemented more easily in cold-atom experiments than the $\mathbb{Z}_2$ LGT~\eqref{eq:z2_hamiltonian}, as it does not require the complete elimination of bare bosonic tunnelling. Both this and the correlated tunnelling term can be obtained as second-order processes in ultracold bosonic mixtures (Fig.~\ref{fig:scheme}{\bf (b)}). In particular, the ratio between these two terms can be controlled experimentally using a Feshbach resonance~\cite{Chanda_2020}, with 
\begin{equation}
\alpha / t = \left(U^{\downarrow}_{\rm ba} - U^{\uparrow}_{\rm ba}\right) / \left(U^{\downarrow}_{\rm ba} + U^{\uparrow}_{\rm ba}\right),
\end{equation}
where $U^{\sigma}_{\rm ba}$ denotes the interspecies Hubbard interactions and $\sigma = \uparrow, \downarrow$ is the internal hyperfine state of one species, simulating the field degrees of freedom.

The $\mathbb{Z}_2$ BHM also resembles the Su-Schrieffer-Heeger (SSH) model~\cite{Su_1979}, with the $\mathbb{Z}_2$ field playing the role of phonons in solid-state systems. The former can thus be understood as simplified lattice degrees of freedom, allowing to simulate a dynamical lattice using ultracold atoms that otherwise are subjected to a static optical potential~\cite{Bloch_2012}. Similarly to the SSH model, the $\mathbb{Z}_2$ BHM also presents Peierls instabilities~\cite{Peierls_1955}, even in the absence of a Fermi surface, giving rise to Bond Order Wave (BOW) phases where the field develop long-range order (Fig.~\ref{fig:scheme}{\bf (c)})~\cite{Gonzalez-Cuadra_2018}. At half filling, the spontaneous symmetry breaking (SSB) of translational invariance gives rise to two degenerate ground states characterized by a dimerized tunnelling pattern. Although both states share the same symmetry properties, they can be distinguished using a topological invariant such as the Berry phase $\gamma$~\cite{Xiao_2010}, quantized to $0$ or $\pi$ in the presence of inversion symmetry. A value of $\gamma = \pi$ corresponds to non-trivial bulk topology, also signalled by the presence of localized states at the system's boundary carrying fractional particle number (Fig.~\ref{fig:scheme}{\bf(c)}).

The topological Peierls insulators with BOW order found in the $\mathbb{Z}_2$ BHM are thus examples of interaction-induced symmetry-breaking topological phases. The interplay between long-range order and non-trivial topology gives rise to strongly-correlated effects that are absent in topological phases where SSB does not take place. For instance, upon doping the system above half filling, the ground state presents topological defects that separate the two symmetry-broken degenerate configurations (Fig.~\ref{fig:scheme}{\bf(d)}). Each of these defects separates regions with different bulk topology, hosting localized fractional states~\cite{Gonzalez-Cuadra_2020a, Gonzalez-Cuadra_2020b}. These are similar to the edge states found at the boundary, but in this case the defect-fractional boson composite quasi-particle can move within the system's bulk, contributing to its transport properties.

\begin{figure}[t]
  \centering
  \includegraphics[width=1.0\linewidth]{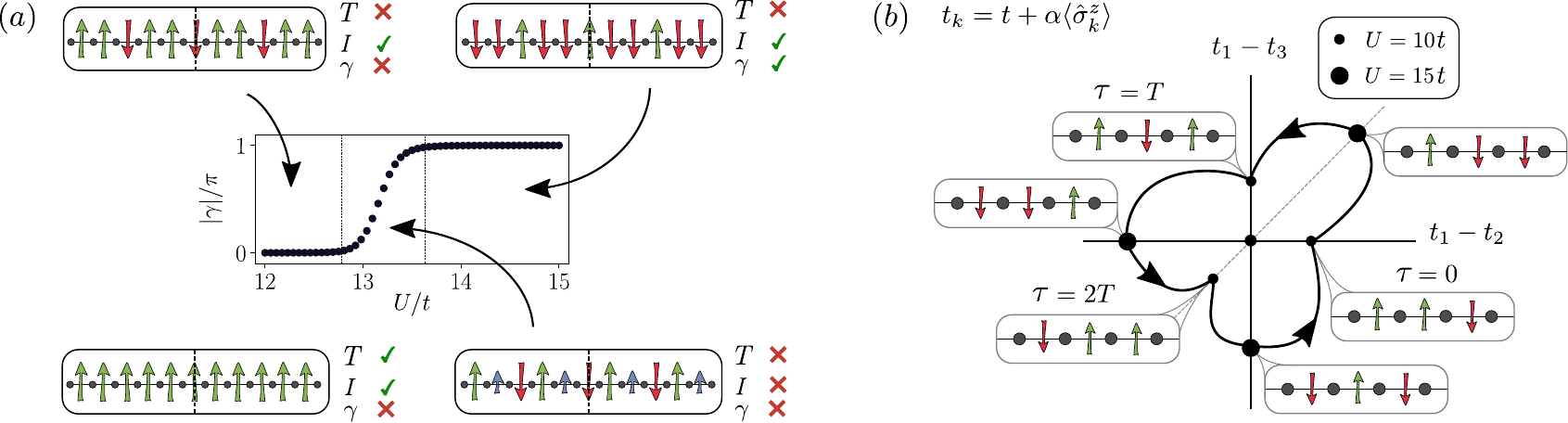}
\caption{\label{fig:emergence} {\bf Emergent symmetry protection and fractional pumping~\cite{Gonzalez-Cuadra_2019b}:} \textbf{(a)} For fractional densities other than half filling, the SSB of translational invariance (T) can also break the protecting inversion symmetry (I). For $\rho = 1 / 3$ and $\rho = 2 / 3$, the latter emerges again for sufficiently strong values of $U$, giving rise to non-trivial topology ($\gamma = \pi$) while maintaining a trimmerized pattern along the phase transition~\cite{Gonzalez-Cuadra_2019b}. \textbf{(b)} Such a symmetry-constrained transition can be employed to devise a self-adjusted pumping protocol, where the system travels adiabatically along the trivial and topological (three-fold) degenerate configurations. Each trivial-topological-trivial subcycle transport $1/3$ of a boson~\cite{Gonzalez-Cuadra_2019b}. In the figure, $\langle \hat{\sigma}^z_k \rangle$, with $k=1,2,3$ denotes the expectation value of the field within the repeating unit cell.}
\end{figure}

Still richer phenomenology can be found at other fractional values of the bosonic density. At $\rho = 1/3$ and $\rho = 2/3$, the Peierls instability gives rise to BOW order with a three-site unit cell (Fig.~\ref{fig:emergence}{\bf(a)}). Among all the possible configurations, only two present inversion symmetry, protecting non-trivial topological properties. Remarkably, such configurations appear in the ground states for certain values of the Hubbard interaction~\cite{Gonzalez-Cuadra_2019b}. We observe, in particular, an interaction-induced topological phase transition where the protecting symmetry is first spontaneously broken in an intermediate phase and then emerges again (Fig.~\ref{fig:emergence}{\bf(a)}). The mechanism behind this process is related to the Peierls constraint: for these values of the bosonic density it is energetically favorable to keep a three-site unit cell. A direct topological transition between the two inversion-symmetric patterns would go through a point where translational invariance is restored. Instead, it is more favorable to break inversion symmetry in an intermediate region to also keep translational invariance broken. 

Such symmetry-constrained topological phase transitions are again a consequence of the interplay between SSB and topology. Furthermore, they can be used to devise a self-adjusted fractional pumping mechanism~\cite{Thouless_1983}. A pumping protocol is a periodic adiabatic modulation where the system's ground state follows a trajectory in parameter space encircling a critical point, transporting in the process a quantized number of particles. If the critical point is topological, this value will be non-zero, and can be computed in terms of a topological invariant. This process normally requires the protecting symmetry to be explicitly broken. In our case, this happens spontaneously just by modifying the Hubbard interactions. Adding an extra field to break the ground state degeneracy and to choose the degenerate configuration to which the system transitions, we can create a protocol that encircles the critical point and go through all degenerate trivial and topological configurations (Fig.~\ref{fig:emergence}{\bf(b)}). The cycle can be divided in three and the particle number transported in each of them is topologically quantized to the fractional value $1/3$.

In this section, we have discussed the topological properties of the $\mathbb{Z}_2$ BHM for some specific densities. Peierls instabilities, however, can occur {\it a priori} for any density. It turns out that this is also the case here, where a staircase of topological Peierls insulators is found in the phase diagram, together with regions where incommensurate orders develop, giving rise to Peierls supersolids~\cite{Chanda_2020}. The above results indicate how ultracold atoms in optical lattices can be used to explored interested strongly-correlated phenomena inspired, but not restricted to, those appearing in condensed matter and high-energy physics.

\section{The synthetic Creutz-Hubbard model }
\label{Emanuele}

\begin{figure}[t]
  \centering
\includegraphics[width=1.0\linewidth]{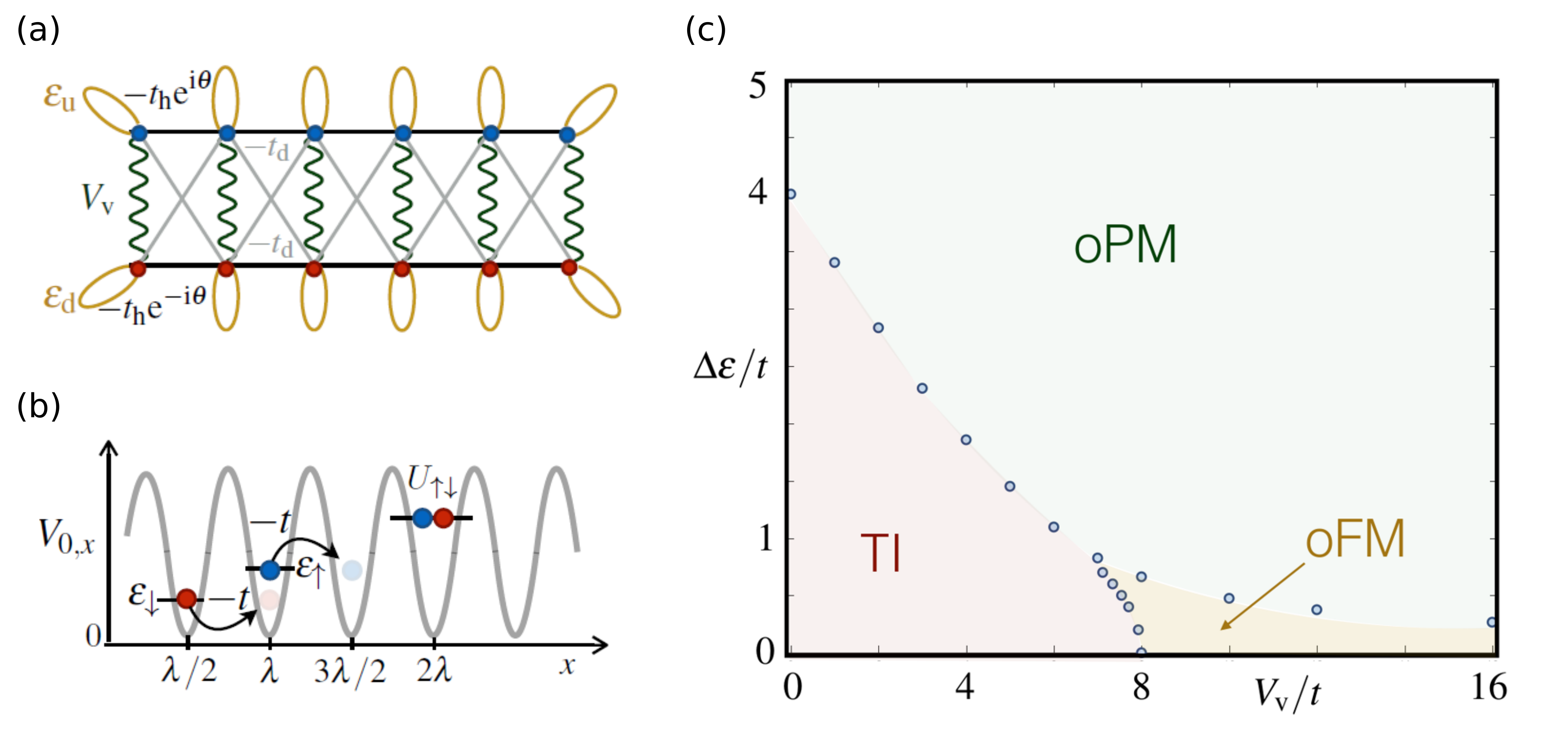}
\caption{\label{fig:chm_sketch} {\bf Creutz topological insulator:} (a) The imbalanced Creutz ladder defined in Eq. (\ref{eq:ch_ham}) in the $\pi$-flux limit. (b) Atoms in two hyperfine states $| \uparrow\rangle$, $| \downarrow\rangle$ are trapped at the minima of an optical lattice. At low temperatures, the kinetic energy of the atoms can be described as a tunnelling of strength $-t$ between the lowest energy levels $\epsilon_{\uparrow}$, $\epsilon_{\downarrow}$ of neighbouring potential wells. Additionally, the s-wave scattering of the atoms leads to contact interactions of strength
$U_{\uparrow \downarrow}$ whenever two fermionic atoms with different internal states meet on the same potential well.
(c) Phase diagram of the C-H model. It displays a topological insulator phase (TI), and two other non-topological phases, namely an orbital phase with long-range ferromagnetic Ising order (oFM), and an orbital paramagnetic phase (oPM). The blue circles label numerical results and the coloured phase boundaries are a guide to the eye. }
\end{figure}

Understanding the robustness of topological phases of matter in the presence of interactions poses a difficult challenge in modern condensed matter, showing interesting connections with high-energy physics. An example lies in the physics of topological insulators, which are insulating phases of matter that are not characterized by local order parameters but, instead, by certain topological invariants \cite{bernevig2013topological}. 
Some of these correlated topological insulators such as \textit{Wilson-Hubbard topological matter} can be described in terms of a relativistic quantum field theory (QFT) of massive Wilson fermions with four-Fermi interactions \cite{wilson1977new}, which originally appeared in the context of lattice gauge theories for elementary particle physics \cite{gattringer2009quantum}.
The precursors of this physics were already discussed for noninteracting models and static gauge fields~\cite{bermudez2010wilson,Mazza_2012}.

\textbf{Creutz-Hubbard model:} Here we will focus on the Creutz topological insulator \cite{creutz1999end} and its connection with the Gross-Neveu-Wilson model \cite{gross1974dynamical}. In particular we will consider the imbalanced Creutz model \cite{JPR16} consisting of spinless fermions on a two-leg ladder. 
These fermions are created and annihilated by $c^{\dagger}_{j, l}$, $c_{j, l}$
where $j \in \lbrace 1,\ldots, N \rbrace$ labels the lattice sites within
the upper or lower legs $l \in \lbrace u,d \rbrace$ and evolve according to 
the tight-binding Hamiltonian
\begin{equation}
H_C=\sum_{j l} \left(-t_l c^{\dagger}_{j+1,l} c_{j,l}-t_x c^{\dagger}_{j+1,l} c_{j,\bar{l}}+\frac{\Delta \epsilon_l}{4} c^{\dagger}_{j,l} c_{j,l} +H.c. \right).
\label{eq:CH}
\end{equation}
Here, $t_l=t e^{-i\pi s_l /2}$ represents the horizontal hopping strength dressed by magnetic $\pi$-flux, $t_x$ stands for the diagonal hopping,
$\Delta \epsilon_l = \Delta \epsilon s_l$ with $\Delta \epsilon >0$ is a energy imbalance between the legs of the ladder, and we use the notation
$s_u=1$ ($s_d=-1$) and $\bar{l}=d$ ($\bar{l}=u$) for $l=u$ ($l=d$).
We now consider the addition of a quartic interaction term to the Hamiltonian
\begin{equation}
H_v=\frac{V_v}{2} \sum_{j,l} c^{\dagger}_{j,l} c^{\dagger}_{j,\bar{l}} c_{j,\bar{l}} c_{j,l},
\end{equation}
and denote the full Hamiltonian as the imbalanced Creutz-Hubbard Hamiltonian
\begin{equation}
\label{eq:ch_ham}
H_{CH}=H_C+H_V, 
\end{equation} 
describing a ladder of strongly-correlated fermions whose tunnelling and interactions are depicted in Fig. \ref{fig:chm_sketch}{\bf(a)}.

The Creutz-Hubbard model could be implemented with ultra-cold fermions in intensity-modulated optical lattices. Moreover, this model represents a  workhorse in the study of strongly correlated topological phases in so-called synthetic quantum matter in atomic, molecular, and optical (AMO) platforms, more particularly, with ultracold gases of neutral atoms in optical lattices (see Fig. \ref{fig:chm_sketch}{\bf(b)}). Indeed, two accessible AMO ingredients, such as as (i) a simple Zeeman shift between the atomic internal states, and (ii) Feshbach resonances, lead to a leg imbalance enabling tuning of the Hubbard type interaction in the ladder. 

The model has an interesting phase diagram  shown in Fig. \ref{fig:chm_sketch}{\bf(c)}. In the non-interacting regime we can rewrite the
Hamiltonian $H_C$ in momentum space as $H_C=\int^{\pi}_{-\pi} \Psi^{\dagger}(k) h_C(k) \Psi(k)$ with the single particle Hamiltonian 
\begin{equation}
\label{eq:single_part_creutz}
h_{\rm C}({k})=-2t_{\rm x}\sigma^x\cos k+\left(\tfrac{1}{2}\Delta\epsilon+2t\sin k\right)\sigma^z.
\end{equation}
This single-particle Hamiltonian respects the sub-lattice symmetry $S$: $U_S h_C(k)U^{\dagger}_S=-h_S(k)$ but breaks both time-reversal and particle-hole symmetries. For this reason the imbalanced Creutz model yields a symmetry-protected topological phase in the AIII class \cite{altland1997nonstandard}.
For $\Delta \epsilon=0$, one finds that the system develops two topological 
flat bands. This flat bands have an associated topological invariant that
can be defined through the Berry connection $\mathcal{A}_{\pm}(k)=i \langle \epsilon_{\pm}(k)| \partial_k |\epsilon_{\pm}(k) \rangle=1/2$. The uniform
Berry connection leads to a finite Zak's phase \cite{zak1989berry} as
$\phi_{Zak,\pm}=\int dq \mathcal{A}_{\pm}(q)$ and equals $\phi_{Zak,\pm}=\pi$. Switching on the leg imbalance $\Delta \epsilon >0$ leads to some curvature in energy bands. The Berry connection becomes non-uniform and depends on the band curvature $\phi_{Zak,\pm}=\pi \theta(f-1)$ (where $f=4t/\Delta \epsilon$ is the curvature). Hence the Zak's phase yields a topological effect until $f>1$ i.e. $\Delta \epsilon < 4t$. For $f<1$ ($\Delta \epsilon > 4t$), the band curvature is large and no topological phenomena occur. This marks a quantum phase transition between the AIII topological insulator and a trivial band insulator as shown along the vertical axis as Fig. \ref{fig:chm_sketch}{\bf(c)}.    
Moreover, turning on interactions leads to a competition between topological phases and two different phases of orbital magnetism. As shown in Fig. \ref{fig:chm_sketch}{\bf(c)}, at large interaction strength a long-range in-plane ferromagnetic order arises, related to the symmetry-broken phase of an orbital quantum Ising model, while the Zeeman imbalance then drives a standard quantum phase transition in the Ising universality class towards an orbital paramagnetic phase.   
 Recently, it has been pointed out that the so-called mean chiral displacement, an observable readily available in after-quench dynamical experiments, could provide a faithful readout of such tripartite phase diagram~\cite{Haller_2020}.
On top of that -- similarly to what was observed for Peierls insulators in Sec.\ref{Daniel} -- a staircase of gapped phases emerges at fractional filling fractions and exhibits (symmetry protected) topological signatures~\cite{Barbarino_2019}.

\textbf{Wilson-Hubbard model:} In the thermodynamic limit,
the rungs of the ladder play the role of the $1$d Bravais lattice 
$j a \rightarrow x \in \Lambda_l=a Z^d=\left\lbrace x: x_i \in Z, \forall i=1,\ldots,d \right\rbrace$, while the ladder index $l\in \lbrace u,d \rbrace$ plays the role of the spinor degrees of freedom of the Fermi field 
$\Psi(x)=\left(c_{j,u},c_{j,d}\right)^t$. Making a gauge transformation 
$c_{j,l}\rightarrow e^{i \pi j/2} c_{j,l}$, one find that the above
imbalanced Creutz model (\ref{eq:CH}) can be rewritten as a \textit{1D Wilson-fermion Hamiltonian} lattice field theory (LFT). 
In general the Wilson LFT is defined by the following Hamiltonian
\begin{align} 
\label{WFH_LFT}
\begin{split}
\mathcal{H}_W = \sum_{x \in \Lambda_l} \big[   \Psi^{\dagger}(x) \left( \frac{i \alpha_i}{2} + \frac{\delta m_i \beta}{2} \right) \Psi(x+a u^i) + m \Psi^{\dagger}(x) \frac{\beta}{2} \Psi(x) \\
+ H.c. \big]  +\sum_{\nu \mu} \sum_{x \in \Lambda_l} \Psi^{\dagger}_{\mu}(x) \Psi^{\dagger}_{\nu}(x) \frac{u_{\mu \nu}}{2} \Psi_{\mu}(x) \Psi_{\nu}(x)  
\end{split}
\end{align} 
where the parameters $\delta m_i$ quantify a certain mass shift introduced to put the fermion doublers up to the cut-off scale  of the LFT, and $u_{\mu,\nu}=(1-\delta_{\mu,\nu}) g^2$ encode the interaction strengths.
In the long-wavelength approximation, the Wilson-fermion LFT yields a continuum Wilson-fermion QFT for $N_D$ instances of the massive Dirac QFT, each describing a relativistic fermion with a different Wilson mass $m_n$ and single-particle Hamiltonian 
$h^n(k)=\alpha^n_i k^i +m_n \beta$. 
In the case of the Creutz model, the Wilson LFT is found by making the following identification of Dirac matrices $\alpha=\sigma^x$, $\beta=\sigma^z$; it describes $N_D=2$ Wilson fermions. 
Moreover, from the perspective of topological insulators, the topological invariants are non-local quantities and thus also sensitive to the Wilson fermion masses. Indeed for the Chern characters $Ch_1$ or Chern-Simons forms $CS_1$ \cite{ryu2010topological} we have that $Ch_1=\frac{1}{2} \sum_n p_n sign(m_n)$ and $CS_1=\frac{1}{4} \sum_n p_n sign(m_n)$ \cite{TRS18}. This means that if the lattice parameters $\delta m_i$ are chosen such that a mass inversion occurs for some of the Wilson fermions, it becomes possible to obtain a non-vanishing integer-valued topological invariant. 
%it yields a continuum Wilson-fermion QFT described by $N_D$ instances of the massive Dirac QFT each describing a relativistic fermion with a different Wilson mass. In the case of the Creutz model, the Wilson LFT is found 
%making the following identification of Dirac matrices $\alpha=\sigma^x$, $\beta=\sigma^z$ and describes $N_D=2$ Wilson fermion. 
%Moreover, from the perspective of topological insulators, the topological invariants are non-local quantities obtained by integrating over all momenta and are thus also sensitive to the masses of the Wilson fermions. Indeed 
%for the Chern characters $Ch_1$ or Chern-Simons forms $CS_1$ \cite{ryu2010topological} we have that $Ch_1=\frac{1}{2} \sum_n p_n sign(m_n)$ and $CS_1=\frac{1}{4} \sum_n p_n sign(m_n)$ \cite{TRS18}. This means that if the lattice parameters $\delta m_i$ are such that a mass inversion occurs for some of the Wilson fermions, it becomes possible to obtain a non-vanishing integer-valued topological invariant $Ch_1$ which can be related to the plateaus of the quantum Hall effect observed at integer fillings. 

\begin{figure}[t]
  \centering
\includegraphics[width=1.0\linewidth]{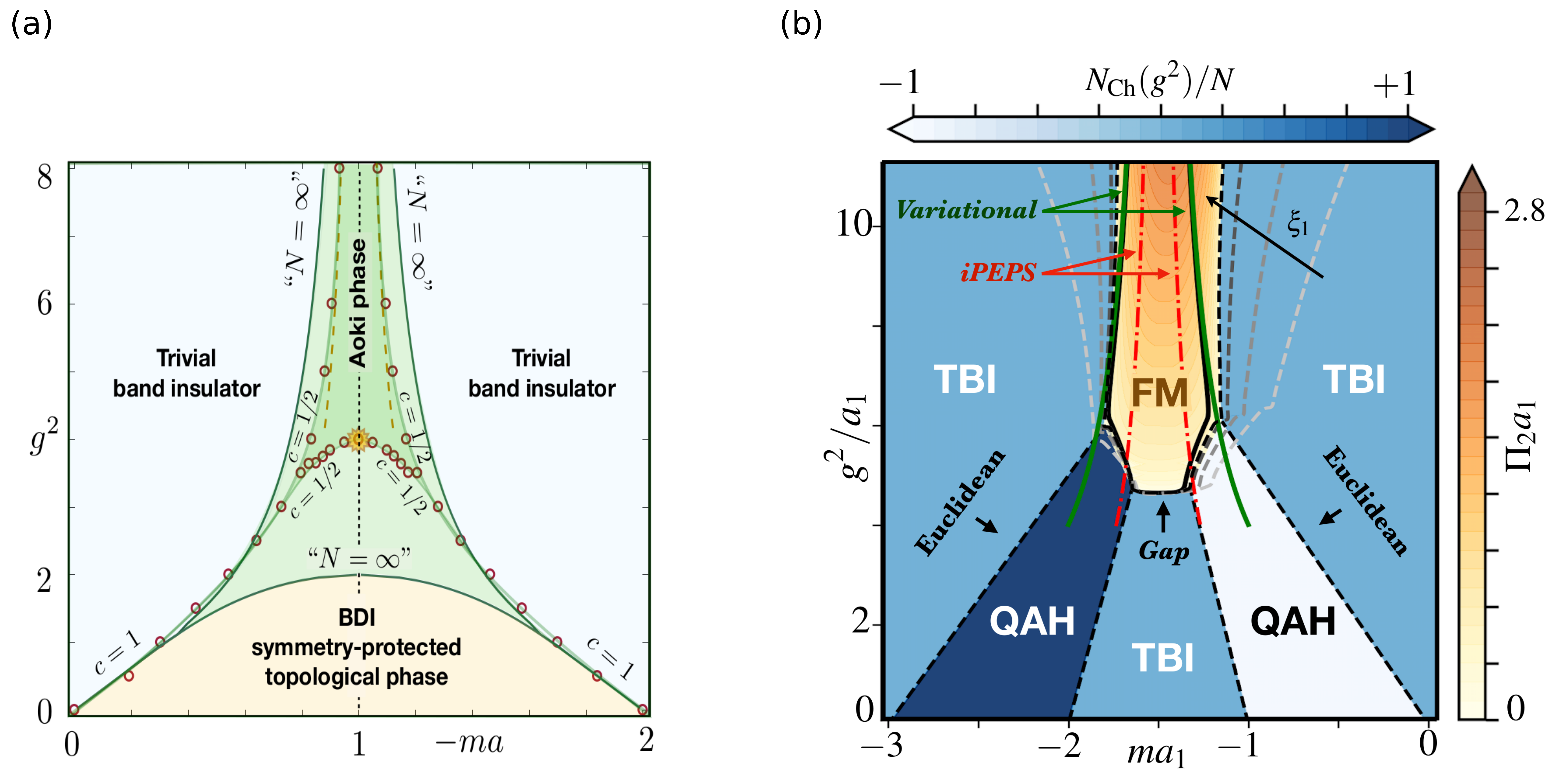}
\caption{\label{fig:gn_phase_diagram} {\bf Gross-Neveu model phase diagrams:} (a) Phase diagram of $(1+1)d$ Gross-Neveu model. The two green solid
lines correspond to the critical lines found by large-$N$ techniques. Red circles represent the critical points of the $N=1$ Gross Neveu lattice model obtained with MPS. The semi-transparent green lines joining these points delimit the trivial band insulator, Aoki phase, and the $\mathsf{BDI}$ symmetry-protected topological phase.
We also include the exact critical point at $(-ma,g^2)=(1,4)$, which is depicted by an orange star, and the strong-coupling critical lines that become exact in the limit of $g^2\to\infty$, depicted by dashed orange lines.  MPS predictions match these exact results remarkably well. 
(b) Phase diagram of $(2+1)d$ Gross-Neveu model. Contour plots of the Chern number $N_{\rm Ch}$ (blue) and $\pi$ condensate (orange),  predicting  large-$N$ QAH phases, trivial band insulators (TBIs), and a ferromagnetic phase (FM). The black solid line is obtained by solving self-consistent equations ({\it Gap}), which can delimit the area of  the FM, but give no further information about the TBI or QAH phases. The green solid lines ({\it Variational}) represent the  product-state prediction for the compass model, and the red dashed-dotted lines correspond to iPEPs. }
\end{figure}

\textbf{Gross-Neveu-Wilson model:} It is possible to  map 
the imbalanced Creutz-Hubbard ladder to a discretized version of the Gross-Neveu field theory: the Gross-Neveu-Wilson field theory. This QFT describes Dirac fermions with $N$ flavors interacting via quartic coupling which live in one spatial and one time dimension. In the continuum, the model is described by the following normal-ordered Hamiltonian $H=\int dx : \mathcal{H} :$ with 
\begin{equation}
    \mathcal{H}= - \bar{\Psi}(x) i \gamma^1 \partial_x \Psi(x) - \frac{g^2}{2N} \left(\bar{\Psi}(x) \Psi(x) \right)^2 .
\end{equation}
Here $\bar{\Psi}(x)=\left( \bar{\Psi}_1(x), \ldots, \bar{\Psi}_N(x)  \right)$ where $\bar{\Psi}_n(x)= \Psi^{\dagger}_n(x) \gamma^0$ are two-component spinor field operators for the n-th fermionic species, and $\gamma^0=\sigma^z$, $\gamma^1=i\sigma^y$ are the gamma matrices, which can be expressed in terms of Pauli matrices for $(1+1)$-dimensional Minkowski space-time, leading to the chiral matrix $\gamma^5=\gamma^0 \gamma^1=\sigma^x$. Therefore, the Gross-Neveu model describes a collection of $N$ copies of massless Dirac field coupled via quartic interactions. 
According to the above discussion and using the exact relations $ma = \frac{\Delta \epsilon }{4 t} -1$ and $g^2=\frac{V_v}{2 t}$, 
the Wilsonian discretization of the Gross-Neveu QFT on a uniform lattice (defined in Eq. \ref{WFH_LFT}) is gauge-equivalent to the imbalanced Creutz-Hubbard model of condensed-matter physics \cite{BTR18}.
Fig. \ref{fig:gn_phase_diagram}{\bf(a)} shows the phase diagram of the
model. We have applied large-$N$ techniques borrowed from high-energy physics, complemented with the study of topological invariants from condensed matter, to unveil a rich phase diagram that contains a wide region hosting a BDI topological insulator. This region extends to appreciable interaction strength $g^2$, and thus corresponds to a strongly-correlated symmetry-protected topological phase.  
Moreover, for sufficiently strong interactions, a gapped phase where parity symmetry is spontaneously broken, viz. the Aoki phase, is formed due to the appearance of a pseudoscalar fermion condensate $\Pi \propto \langle \bar{\Psi} i \gamma^5 \Psi \rangle$. The large-$N$ prediction has allowed us to find the critical line separating the topological insulator from the Aoki phase by studying the onset of the pseudoscalar condensate, and show that it terminates at tricritical point where all
three phases of matter coexist.     
%\textbf{$(2+1)$d Gross-Neveu-Wilson model:}

The model can easily be extended to $(2+1)$ dimensions \cite{ziegler2020correlated}. In Fig. \ref{fig:gn_phase_diagram}{\bf(b)} we show the complete phase diagram of the model together with the contour plot of the Chern number and the pseudo-scalar condensate. 
In the non-interacting regime $g^2\approx0$, the single-flavor $(2+1)$d Gross-Neveu-Wilson model corresponds to the square-lattice version \cite{qi2006topological} of the Haldane model \cite{haldane1988model}
of the quantum anomalous Hall effect (QAH). This model has a quantized  
Hall conductance $\sigma_{xy}=\frac{e^2}{h}N_{\rm Ch}$ with $ N_{\rm Ch}=\frac{N}{2}\sum_{\boldsymbol{n}_d}(-1)^{(n_{d,1}+n_{d,2})}{\rm sign}(m_{\boldsymbol{n}_d})$. With the interaction switched on the fermions
can also form polar condensates $\Pi_1 \propto \langle \bar{\Psi}\gamma^1 \Psi\rangle$ and $\Pi_2 \propto \langle \bar{\Psi}\gamma^2 \Psi\rangle$ related to the spontaneous breaking
of parity symmetry and Lorentz-invariance. The set of parameters
where the $\Pi$ condensates form define critical line separating 
the correlated QAH phase from long-range-ordered ferromagnets as depicted in Fig. \ref{fig:gn_phase_diagram}{\bf(b)}.  

\section{Experimental and experimentally friendly quantum simulators \label{Valentin}}
Spectacular advances in the field of artificial quantum systems make it possible to realize lattice gauge theories with quantum simulators in table-top experiments. Platforms that succeeded in achieving this goal so far are trapped ions~\cite{martinez2016real, kokail2018self}, ultracold atoms~\cite{Goerg2019, Schweizer2019, Mil2020, Yang2020}, and superconducting qubits~\cite{Satzinger2021}. A common characteristic shared by these experimental platforms is their high tunability and precise readout, %In this way, artificial quantum systems offer
which provides a toolbox distinct from those of  particle physics or solid-state experiments. For example, quantum simulators allow for a tuning of the microscopic parameters, studying isolated quantum many-particle dynamics in real time, or measuring higher-order correlations~\cite{Endres2011,Schweigler2017} and entanglement~\cite{Islam2015}. 

In the following, we briefly review four quantum simulation experiments of Abelian lattice gauge theories based on ultracold atomic systems and then discuss recent developments towards experimentally friendly implementations. %proposals for quantum simulation experiments.

{\it{Density-dependent gauge fields}.} 
%One characteristics 
A defining feature of lattice gauge theories is that the transport of matter is %determined by the 
tied to interactions with the gauge field. Therefore, a first step towards the simulation of genuine lattice gauge theories consists of implementing a coupling mechanism between the gauge and matter fields, which reproduces this characteristic. In the %Zurich
experiment~\cite{Goerg2019} ultracold fermions in an optical lattice were manipulated such that site-occupation dependent Peierls phases mediated the hopping. Since the Peierls phase depends on the density %one calls this construction typically 
this mechanism has been coined density-dependent gauge fields~\cite{Greschner2014}. In order to realize such %Hamiltonian 
dynamics the Fermi-Hubbard Hamiltonian is driven with an inertial force $\hat{V}(\tau)=-f(\tau) \sum_{j, \sigma} j \hat{n}_{j \sigma}$, where $f(t)$ is sinusoidally modulated at two frequencies and $\hat{n}_{j \sigma}$ is the density per site of a given hyperfine state $\sigma$.
This Floquet scheme relies on breaking time-reversal symmetry by driving the lattice simultaneously at two frequencies %and that are
resonant with the on-site interactions. 
For sufficeintly fast driving, one obtains %This induces 
effective density-assisted tunnelling processes that are controllable in amplitude and phase. 
In the experiment~\cite{Goerg2019}, the authors characterized the tunnel coupling and detected two distinct regimes as a function of the modulation amplitudes which can be characterized %these regions 
by a $\mathbb{Z}_2$-invariant. 

{\it{Floquet engineering of a $\mathbb{Z}_2$ lattice gauge}.} 
The experimentalists of~\cite{Schweizer2019} 
%went 
also used the Floquet approach to realise a minimal instance of a $\mathbb{Z}_2$ lattice gauge theory. %To this end, 
They achieved this by manipulating two components, $|a\rangle \equiv\left|F=1, m_{F}=-1\right\rangle$ and $|f\rangle \equiv \left|F=1, m_{F}=+1\right\rangle$, of $^{87}$Rb in a double-well potential with a periodic drive. 
%The two components were $|a\rangle \equiv\left|F=1, m_{F}=-1\right\rangle$ and $|f\rangle \equiv \left|F=1, m_{F}=+1\right\rangle$. 
The double well potential was chosen species-dependent and allowed for both tunnel coupling between neighboring sites and tuning of an energy offset. The energy offset was realized with a magnetic field gradient, making use of the magnetic moment of the Zeeman states. In particular, the offset is only experienced by the $f$-particles. In the limit of strong on-site interactions, the direct tunnelling processes are suppressed but can be restored through a periodic modulation at the resonance frequency. 
In the cases of resonant periodic driving, i.e. at the on-site interaction strength and judiciously chosen modulation parameters, the effective Floquet Hamiltonian turns into the desired two-site $\mathbb{Z}_2$ lattice gauge theory~\cite{BSA18}. Here, matter and gauge fields are represented by two different species, %which in turn are realized by 
corresponding to two Zeeman levels of the hyperfine groundstate manifold of $^{87}$Rb. Matter fields is associated with $a$-states and the  gauge field by the number imbalance of the $f$-states. 

{\it{Local U(1) symmetry from spin-changing collisions}.}
In the experimental work~\cite{Mil2020}, a scalable analogue quantum simulator of a U(1) gauge theory was realized. Specifically, the experiment employed a mixture of two atomic Bose-Einstein condensates, $^{23}$Na and $^{7}$Li. The atoms in the two hyperfine states of sodium form an effective large spin representing the gauge fields, while the two hyperfine states of lithium correspond to two lattice sites of bosonic matter. The gauge invariant interactions between matter and gauge fields were engineered using inter-species spin changing collisions based on proposals of~\cite{zohar2013quantum, kasper2016schwinger, kasper2017implementing, zache2018quantum}. As an application, particle production from vacuum in the presence of a strong gauge field has been quantum-simulated. 

{\it{Emergent local U(1) symmetry for bosons in a superlattice}.} 
The experiment~\cite{Yang2020} %is the first 
realized a large scale quantum simulation of a U(1) lattice gauge theory on 71 sites and has thus made a major step towards quantum simulation in the thermodynamic limit. %For the quantum simulation,
A strongly tunable Bose-Hubbard system, i.e. ultracold atoms in an optical superlattice, was employed. The gauge invariant interactions between matter and gauge field arise from %the correct
a suitable choice of Hubbard parameters, which effectively suppresses unwanted processes via an energy penalty~\cite{fradkin2013field,banerjee2012atomic}. In the limit of strong interactions, this atomic system can then be rigorously mapped to a U(1) lattice gauge theory with fermionic matter. %In this experiment, we 
The authors not only realized a U(1) lattice gauge theory, but also quantified how well gauge invariance is satisfied in this quantum simulation. The Gauss law has been evaluated by measurements of lattice site occupation and density-density correlations, which allow to quantify the local constraints imposed by gauge invariance. Furthermore, the large degree of tunability of this setup enabled a scan across a quantum phase transition. 

\subsubsection*{Challenges and new directions}
These four experiments illustrate that the quantum simulation of Abelian lattice gauge theories with ultracold atom systems is a reality. Having passed this milestone, new experimental challenges are now being addressed. %The following directions are receiving a large amount of attention: 
In particular, there is a continuing community effort in the following directions: 
(i) lattice gauge theories in higher dimensions~\cite{Wiese2013, zohar2015quantum}, (ii) non-Abelian symmetries~\cite{Zohar2013nonAbelian,banerjee2013atomic, tagliacozzo2013simulation}, (iii) lattice gauge theories with bosonic matter (see Sec.~\ref{Titas}), and (iv) the effects of violating the local symmetries~\cite{Halimeh2020, Halimeh2020a}.

The quantum simulation of lattice gauge theories in higher dimensions has already achieved its first successes. For example, two-dimensional gauge theories have been realized in Rydberg arrays~\cite{Semeghini2021} or superconducting qubits~\cite{Satzinger2021}. Ultracold atom lattice experiments have realized a single ring-exchange interaction~\cite{Dai2017}, but no large-scale quantum simulation yet. Another challenge is the quantum simulation of non-Abelian lattice gauge theories, where first attempts of digital simulations on universal quantum computers exist. However, the analogue quantum simulation of non-Abelian theories is still in its infancy. Many initial proposals for analogue quantum simulations of non-Abelian theories are beyond the current capabilities of ultracold atomic systems. Hence, there is great interest in experimentally feasible proposals for non-Abelian lattice gauge theories.

{\it{Minimal SU(2) models for ultracold atom systems}.}
One outstanding challenge confronting the quantum simulation of non-Abelian lattice gauge theories is that one has to obey not one, but several non-commuting, local constraints~\cite{Kogut1975}. However, instead of enforcing the local symmetry one can also eliminate the gauge fields or matter fields and quantum simulate the resulting reduced systems~\cite{muschik2017u}. This encoding strategy was already successfully used for the first quantum simulation of the lattice Schwinger model with an ion quantum computer~\cite{martinez2016real}. The encoding strategy can be applied for one-dimensional systems where the Gauss law permits elimination of gauge degrees of freedom. Hence, one expects that this strategy can also work for non-Abelian lattice gauge theories. Moreover, the encoding approach can also be used for computational purposes, e.g. in the context of MPS~\cite{Banuls2017}. In \cite{Kasper2020b} the encoding strategy was used for two-site models to propose a quantum simulator of an SU(2) lattice gauge theory and was mapped on a three- and four-level system. The four-level model can then be realized, e.g., with two coupled two-level systems like Rydberg atoms or coupling hyperfine states of atoms via microwave radiation. This mapping allows the study of certain minimal non-Abelian lattice gauge theories with current quantum simulator technology.

{\it{Non-Abelian gauge invariance from dynamical decoupling}.}
Next we discuss a scalable strategy to quantum simulate non-Abelian lattice gauge theories. The earliest proposals for implementing non-Abelian lattice gauge theories with ultracold atom systems used atomic collisions~\cite{zohar2013quantum}, complex energy penalties~\cite{zohar2011confinement, banerjee2012atomic}, or elaborate digital schemes~\cite{Zohar2017Digital}. %One of the most 
Another promising approach is the engineering of dissipation to enforce non-Abelian gauge invariance~\cite{stannigel2014constrained}. %The dissipation proposal 
This dissipative approach is elegant since it  requires a linear coupling to the Gauss law operators, but needs a correctly engineered dissipation. An immediate question is whether one can also enforce non-Abelian local symmetry through a coherent drive.

The question of enforcing global symmetries via coherent driving was previously addressed in the field of quantum information. In %quantum information 
this context, %one wants to protect 
the aim is to protect a target (sub)space for information processing. One strategy to minimize the interaction with the environment is dynamical decoupling \cite{Lidar2013}. In the first work~\cite{Viola1999} periodic bang-bang protocols were used to enforce global symmetries. In~\cite{Kasper2020a}, the authors generalized this idea to a periodic bang-bang control to enforce a {\it local} symmetry. As a consequence the gauge invariant part of the Hilbert space decoupled from undesired sectors. The concept was demonstrated for a one dimensional SU(2) lattice gauge system, but generalization to other non-Abelian symmetries and more dimensions is straightforward. A major advantage of the coherent driving approach %has the great benefit for not only being
is its generality. The application to ultracold atom systems is particularly appealing due to the local nature of typical gauge-violating processes, but the general scheme remains applicable for other platforms and thus opens a promising route for the quantum simulation of non-Abelian lattice gauge theories.

{\it{Rotor Jackiw-Rebbi model}.}
Efforts towards simplifying the quantum simulation of lattice gauge theories also raise theoretical questions: %Do we need to quantum simulate 
which properties of lattice gauge theories are crucial to study high-energy systems or strongly correlated solids? Do simpler models without gauge invariance exhibit similar physics as gauge theories? %Motivated by these question. 
To address these questions, the work~\cite{Gonzalez-Cuadra2020} introduced a rotor-regularized version of the Jackiw-Rebbi model in (1+1)-dimensions. The original Jackiw-Rebbi quantum field theory~\cite{Jackiw1976} describes Dirac fermions interacting with a bosonic field. In contrast, the authors of~\cite{Gonzalez-Cuadra2020} studied a regularized version of the Jackiw-Rebbi: fermions $c_i$ on a lattice interacting with spins $\mathbf{S}_i$ on each lattice site. The regularized Hamiltonian is 
\begin{equation}
H=\sum_{i}\left[-t\left(c_{i}^{\dagger} c_{i+1}+c_{i+1}^{\dagger} c_{i}\right)+\mathbf{g} \cdot \mathbf{S}_{i} c_{i}^{\dagger} c_{i}-\mathbf{h} \cdot \mathbf{S}_{i}\right] \,.
\end{equation}
This model admits a simple implementation with an ultracold mixture in an optical lattice. Further, extensive numerical/analytical studies~\cite{Gonzalez-Cuadra2020} revealed the spontaneous breaking of chiral symmetry~\cite{Nambu1961}, as well as a confinement-deconfinement transition. For a wide, experimentally relevant parameter regime, Dirac fermions acquire a dynamical mass via the spontaneous breakdown of chiral symmetry. %This work also studies the effect of chiral symmetry restoration from thermal fluctuations and a confinement-deconfinement quantum phase transition, where mesonlike fermions fractionalize into quarklike quasiparticles bound to topological solitons of the rotor field.
At the confinement-deconfinement quantum phase transition meson-like bound states fractionalize into quark-like quasiparticles bound to topological solitons of the rotor field.

\section{Conclusions and Outlook}
%{\color{red} Matteo}}
\label{Matteo}

In this short essay, we have offered an overview of the rapidly developing research field of analogue quantum simulations for lattice gauge theory models. 
Since it is arduous to recapitulate the many significant achievements of the last decade in such a brief text, we have focused almost exclusively on contributions of collaborations involving the Quantum Optics Theory group at ICFO, with no intention of assessing their relevance against the work of others.

The most acknowledged goal of analogue quantum simulations is to create platforms for the embodiment of quantum many-body problems, which are both more tunable in their couplings and more accessible in their observables than their original counterparts in nature. 
This should ultimately serve to address open questions in a more decisive way than traditional computational techniques: something which often is dubbed as quantum supremacy, and that reflects the early vision of R. Feynman.
In this respect, we are now contemplating an exciting explosion of experimental results that start to incarnate a decade of theoretical proposals. We are just starting to scratch an incredible world with a wealth of possibilities ahead.
We could expect that also the first analogue quantum simulators of non-Abelian lattice gauge theories will, in another five years or so, come to experimental birth. 
Such a step will be instrumental in embarking for the as yet unknown, e.g., exploring dynamical phenomena,  or finite densities of matter, currently inaccessible to the best classical computational techniques. 
The first milestones in this respect have been set recently. % thinking about all the Heidelberg-Wien stuff on non-equilibrium phase transitions, for example -- not sure in which respect to cite them here...

There is, however, still another prospect within reach of analogue quantum simulations, namely that of bringing abstract models into real laboratory life.
By this we mean the realisation of theoretical models that have been up to now only blackboard exercises, not directly observed in nature, as we have illustrated here in the case of bosonic matter. 
This will surely generate radically new phenomenology, which is not only exciting for pure curiosity-driven interests but could also  eventually refertilise the field of condensed-matter, with possible technological outcomes on the long run. 
The intimate relation between these two big intellectual constructs, lattice gauge theory and condensed-matter physics, have indeed already shown its potential several times in the past.

\enlargethispage{20pt}

\acknowledgments
This research was supported by 
 National Science Centre (Poland) under project  2017/25/Z/ST2/03029 (T.C., J. Z.).L.T. acknowledges support from the Ram\'on y Cajal program RYC-2016-20594, the ``Plan Nacional Generaci\'on de Conocimiento'' PGC2018-095862-B-C22 and the State Agency for Research of the Spanish Ministry of Science and Innovation through the “Unit of Excellence María de Maeztu 2020-2023” award to the Institute of Cosmos Sciences (CEX2019-000918-M), and by STFC Consolidated Grant ST/T000813/1 (S.H.) . ICFO group acknowledges support from ERC AdG NOQIA, State Research Agency AEI (“Severo Ochoa” Center of Excellence CEX2019-000910-S, Plan National FIDEUA PID2019-106901GB-I00/10.13039 / 501100011033, FPI, QUANTERA MAQS PCI2019-111828-2 / 10.13039/501100011033), Fundació Privada Cellex, Fundació Mir-Puig, Generalitat de Catalunya (AGAUR Grant No. 2017 SGR 1341, CERCA program, QuantumCAT \ U16-011424, co-funded by ERDF Operational Program of Catalonia 2014-2020), EU Horizon 2020 FET-OPEN OPTOLogic (Grant No 899794), and the National Science Centre, Poland (Symfonia Grant No. 2016/20/W/ST4/00314), Marie Sk\l odowska-Curie grant STRETCH No 101029393.  The TDVP simulations have been performed using ITensor library (\url{https://itensor.org}) and with a code based on a flexible Abelian Symmetric Tensor Networks Library, developed by the group of M. R. in collaboration with the group of S. Montangero (Padua). Support of PL Grid infrastructure is acknowledged.

%\disclaimer{Insert disclaimer text here if applicable.}

%%%%%%%%%% Insert bibliography here %%%%%%%%%%%%%%
%\bibliographystyle{ieeetr}

%\bibliography{reviewbib}

\end{document}